\documentclass[fleqn,twoside]{article}
\usepackage{gc,epsf}
\usepackage{graphics}
\heads{S.M.Kozyrev, S. V. Sushkov}
      {Composite wormholes in vacuum Jordan-Brans-Dicke theory.}

\begin{document}
\twocolumn[

\Title{Composite wormholes in vacuum Jordan-Brans-Dicke theory.}

\Author{Sergey M.Kozyrev\foom 1, Sergey V. Sushkov\foom 2}
       {1. Scientific center gravity wave studies ''Dulkyn'',
Kazan, Russian Federation.

        2. Department of General Relativity and
Gravitation, Kazan State University, Kremlevskaya str. 18, Kazan
420008, Russia and Department of Mathematics, Tatar State
University of Humanities and Education, Tatarstan str. 2, Kazan
420021, Russia}

\Abstract
    {New classes composite vacuum wormhole solutions of Jordan-Brans-Dicke gravitation is presented and analysed. It
    is
    shown that such solution holds for both, a bridge between
    separated Schwarzschild and Brans Universes and for a bridge
    connecting two Schwarzschild asymptotically flat regions joined
    by Brans throat. We have also noticed that there are some new
    possible candidates for wormhole spacetimes.}


]

\email 1 {Sergey@tnpko.ru}
\email 2 {sergey.sushkov@ksu.ru}

\section{Introduction}

Lately there has been renewed interest in the scalar-tensor
theories of gravitation. Important arena where these theories have
found immense applications is the field of wormhole physics.
Several classes of solutions in the scalar-tensor theories support
wormhole geometry. The most prominent example of scalar-tensor
theories is perhaps the Jordan-Brans-Dicke (JBD) \cite{Jordan},
\cite{Brans}. Now a day, predictions of JBD theory to be
consistent not only with the weak field solar system tests but
also with the recent cosmological observations.

    JBD theory describes gravitation through a metric tensor g$_{\mu \nu }$ and a massless
scalar field $\phi $. In this theory, static wormhole solutions
were found in vacuum, the source of gravity being the scalar
field. Several static wormhole solutions in JBD theory have been
widely investigated in the literature \cite{Agnese}, \cite{Nandi}.
It was shown that three of the four Brans classes of vacuum
solutions admit a wormholelike spacetime for convenient choices of
their parameters.

The Birkhoff's theorem does not hold in the presence of JBD scalar
field, hence some static JBD solutions seem possible in
spherically symmetric vacuum case \cite{Hanlon}. In this context,
we recall the well known fact that the first exact solutions of
JBD field equations in widespread Hilbert coordinates were
obtained in parametric form by Heckmann \cite{Heckmann}, soon
after Jordan proposed scalar-tensor theory. Apart from this, he
showed that the Schwarzschild solution represent solution of JBD
field equations, too.

In this work, we have borrowed from astrophysicists the idea of
boson stars that is generated by a complex scalar field matching
with a vacuum exterior solution. Furthermore, there is now a
growing consensus that wormholes are in the same chain of stars
and black holes. On the other hand, it is known that the JBD
scalar field $\phi $ plays the role of classical exotic matter
required for the construction of traversable Lorentzian wormholes
\cite{Nandi2}.

This work is organized as follows. After giving a short account of
the JBD theory and its static spherically symmetric solutions, we
explore the wormhole nature of Brans solutions in section 2. In
section 3 composite wormhole nature of the Brans solutions will be
discussed by impose certain junction conditions. In section 4 we
give some new possible candidates for wormhole spacetimes.
Finally, conclusions are drawn in section 5.

\section{Interior solution}

Scalar-tensor theories are described by the following action in
the Jordan frame is:

\begin{eqnarray}
S &=&\int d x\sqrt{-g}(\phi R-\omega \left( \phi \right) g^{\mu
\nu }\nabla
_\mu \phi \nabla _\nu \phi -  \nonumber \\
&&\ \ \;\ \;\ \;\ \;\ \;\ \;\ \;\ \;\ \;\ \;\ \;\ -\lambda \left(
\phi \right) )+S_m.  \label{eq3}
\end{eqnarray}
Here, \textit{R} is the Ricci scalar curvature with respect to the
space-time metric g$_{\mu \nu }$ and S$_m$ denote action of matter
fields. We use units in which gravitational constant \textit{G}=1
and speed of light c=1.
The dynamics of the scalar field $\phi $ depends on the functions $\omega $($%
\phi $) and $\lambda $($\phi $). It should be mentioned that the
different choices of such functions give different scalar-tensor
theories. We restrict
our discussion to the JBD theory which characterized by the functions $%
\lambda $($\phi $) = 0 and $\omega $($\phi $) = $\omega $/$\phi $, where $%
\omega $ is a constant.

Variation of (\ref{eq3}) with respect to g$_{\mu \nu }$ and $\phi
$ gives, respectively, the D-dimensional field equations:

\begin{equation}
R_{\mu \nu }-\frac 12Rg_{\mu \nu }=\frac 1 {\phi \ }T_{\mu \nu
}^M+ T_{\mu \nu }^{JBD},  \label{eq4}
\end{equation}
where

\begin{eqnarray}
T_{\mu \nu }^{JBD} &=& [\frac \omega {\phi ^2}\left( \nabla _\mu
\phi \nabla _\nu \phi -\frac 12g_{\mu \nu }\nabla _\alpha \phi
\nabla
^\alpha \phi \right) +  \nonumber \\[0.01in]
&&\ \ +\frac 1\phi \left( \nabla _\mu \nabla _\nu \phi -g_{\mu \nu
}\nabla _\alpha \nabla ^\alpha \phi \right) ].  \label{eq5}
\end{eqnarray}
and

\begin{equation}
\nabla _\alpha \nabla ^\alpha \phi =\frac{T_\lambda ^{M\ \lambda
}}{3 +2 \omega },  \label{eq6}
\end{equation}
and $T_\lambda ^{M\ \lambda }$ is the energy momentum tensor of
ordinary
matter which obeys the conservation equation $T_{\mu \nu ;\lambda }^{M\ }$ g$%
^{\nu \lambda }$= 0.

Consider spherically symmetric spacetime geometry. The most common
form of line element of a spherically symmetric spacetime in
comoving coordinates can be written as

\begin{eqnarray}
ds^2 &=&-g_{tt}\left( r,t\right) dt^2+g_{rr}\left( r,t\right)
dr^2+
\nonumber  \label{eq1} \\
&&+\mathcal{\rho }^2\left( r,t\right) d\Omega ^2. \label{eq1}
\end{eqnarray}
where d$\Omega ^2$ is the element of solid angle. Four classes of
static JBD theory solutions were derived by Brans himself way back
in 1962. The Brans class I solution (in isotropic coordinates with
$g_{rr} = e^{2\lambda }$, $g_{tt} = e^{2\nu }$, $\rho
^2=r^2g_{rr}$) is given by

\begin{eqnarray}
\phi &=& \phi _0\left( \frac{1-\frac Br}{1+\frac Br}\right)
^{\frac C A },
  \label{eq21} \\
\lambda &=& \lambda _0+\ln \left[ \left( 1+\frac Br\right) ^2\left( \frac{%
1-\frac Br}{1+\frac Br}\right) ^{\frac{A -C-1}A }\right] ,
\nonumber \\
\nu &=& \nu _0+\ln \left[ \left( \frac{1-\frac Br}{1+\frac
Br}\right) ^{\frac 1 A }\right] ,  \nonumber
\end{eqnarray}

where:

\begin{eqnarray*}
A =\sqrt{\left( C+1\right) ^2-C\left( 1-\frac{\omega \ C}2\right)
}, \label{2.9}
\end{eqnarray*}

In order to investigate whether a given solution represents a
wormhole geometry, it is convenient to cast the metric into
Morris-Thorne canonical form:

\begin{eqnarray}
&& ds^2=-e^{2\chi \left( \stackrel{*}{R}\right) }dt^2+\left[
1-\frac{b\left( \stackrel{*}{R}\right) }{\stackrel{*}{R}}\right]
^{-1}dr^2+ \nonumber \\
&& \ \ \ \ \ \ \ \  +\stackrel{*}{R}^2d\Omega ^2,\label{eq111}
\end{eqnarray}
where $\chi \left( \stackrel{*}{R}\right) $ and \textit{b}$\left(
\stackrel{*}{R}\right) $ are arbitrary functions of the radial
coordinate, $\stackrel{*}{R}$. $\chi \left( \stackrel{*}{R}\right)
$ is denoted as the redshift function, for it is related to the
gravitational redshift; \textit{b}$\left( \stackrel{*}{R}\right) $
is called the form function, because as can be shown by embedding
diagrams, it determines the shape of the wormhole \cite{Visser}.
The radial coordinate has a range that increases from a minimum
value at $\stackrel{*}{R}_0$, corresponding to the wormhole
throat, to \textit{a}, where the interior spacetime will be joined
to an exterior vacuum solution. In the case of JBD theory the
expressions for wormhole geometry can be easily obtained by
connecting two Schwarzschild asymptotically flat regions joined by
means of Brans throat. Moreover, one can use even not
asymptotically flat JBD solution for a throat region. The Brans
class I solution can be cast to the form (\ref{eq111}) by defining
a radial coordinate $\stackrel{*}{R}$ which is related with
\textit{r} via the expression
\begin{equation}
\stackrel{*}{R}=r\left( 1+\frac Br\right) ^{1+\frac{C+1}A }\left(
1-\frac Br\right) ^{1-\frac{C+1}A }
\end{equation}
The functions $\chi \left( \stackrel{*}{R}\right) $ and
\textit{b}$\left( \stackrel{*}{R}\right) $ are the given by
\cite{Agnese}
\begin{equation}
\chi \left( \stackrel{*}{R}\right) =\frac 1 A \ln \left[ \left(
\frac{1-\frac Br}{1+\frac Br}\right) \right] ,
\end{equation}
\begin{eqnarray}
b\left( \stackrel{*}{R}\right) =\stackrel{*}{R}\left[ \left(
1-\frac{r^2\left( \stackrel{*}{R}\right) -B^2-2r\left(
\stackrel{*}{R}\right) \frac{B\left( C+1\right) }A }{r^2\left(
\stackrel{*}{R}\right) -B^2}\right) ^2\right]
\end{eqnarray}
The axially symmetric embedded surface $z=z \left(
\stackrel{*}{R}\right) $ shaping the wormhole's spatial geometry
is obtained from
\begin{equation}
\frac{dz}{d\stackrel{*}{R}}=\pm \left[ \frac
{\stackrel{*}{R}}{b\left( \stackrel{*}{R}\right) }-1\right]
^{-\frac 12}
\end{equation}

By definition of wormhole at throat its embedded surface is
vertical. The throat of the wormhole occurs at $\ \stackrel{*}{R}
= \stackrel{*}{R}_0$ such that $\ b\left( \stackrel{*}{R}_0\right)
= \stackrel{*}{R}_0$. This gives minimum allowed \textit{r} -
coordinate radii r$_0^{\pm }$ as
\begin{equation}
r_0^{\pm }=\frac B A \left[ C+1\pm \sqrt{\left( C+1\right) ^2-A
^2}\right]
\end{equation}

According to pioneering Brans and Dicke work  \cite{Brans} to
discuss the perihelion rotation of a planet requires a
specifications of the arbitrary constants in solution (\ref{eq21})
in such a way that this solution agrees in the weak - field limit
as viewed by a distant observer. Hence comparing the expression
for scalar field of Brans class I solution with weak - field
solution we get \cite{Bhadra}
\begin{eqnarray}
C=-\frac 1{\omega +2} \label{eq7}
\end{eqnarray}

On the contrary, one can explain exterior region of JBD
spherically symmetric configurations by the Schwarzschild metric.
In this case we make the reasonable demand that this solution of
scalar-tensor theory field equations lead to free estimates of
arbitrary constants and lower limit of $\omega $. In this context,
we recall the fact that the above conjecture can be easily adopted
in other three classes of Brans solutions.

\section{Junction conditions}

Hawking's theorem \cite{Hawking} in JBD states that the only
spherically symmetric vacuum solution is static and given (up to
coordinate freedom) by the Schwarzschild metric. Using the key
assumption that the Brans class I solution physically acceptable
we shall consider that the JBD scalar field is distributed from
the throat to a radius \textit {a}, where the solution is matched
to an exterior Schwarzschild vacuum spacetime.

\begin{eqnarray}
&& \phi  = 1, \nonumber  \label{eq22} \\
&& \lambda  = ln \left[ \left( 1+\frac \mu r\right) ^2\right] ,
\label{eq22} \\
&& \nu  = \ln \left( \frac{1-\frac \mu r}{1+\frac \mu r}\right),
\nonumber
\end{eqnarray}

In this case the boundary surface entails via the field equations
a jump in second derivations of metric coefficient, but first
derivatives remains continuous so can be used to match to the
vacuum solution. Now in order to justify calling the geometry a
wormhole we need an explicit definition for the constants in Brans
I solution (\ref{eq21}). We have five equations for five unknowns
B, C, $\phi _0$ $\lambda _0$ $\nu _0$. Thus, from the junction
conditions, the interior metric and scalar field parameters can be
determined at the boundary surface in terms of the exterior metric
and scalar field parameters $\mu $, $\phi $ = 1. Hence,
Darmois-Israel junction conditions are fulfilled. The brief
computation yields \cite {Kozyrev}:

\begin{eqnarray*}
\ B &=& a\sqrt{\frac{2a^2-2a\mu +\mu ^2\left( 2+\omega \right) }{
-2a\mu +2\mu ^2+a^2\left( 2+\omega \right) }} , \ \ \ \ \ \ \ \ \ \ \ \ \ \ \ \nonumber \\
\ C &=& \frac{2\left( a^2-a\mu +\mu ^2\right) }{a\mu \
\omega } , \ \ \ \ \ \ \ \ \ \ \ \ \ \ \ \ \ \ \ \ \ \ \ \ \ \ \ \ \ \ \ \ \nonumber \\
\lambda _0 &=& \ln \left( \frac{\left( 1-\frac{2B}{B+a}\right)
^{\frac{ 2a^2+2\mu ^2+a\mu \left( \omega -2\right) }{A \ a\mu \
\omega
}}\left( a+\mu \right) ^2}{a^2-B^2}\right) ,\ \ \ \ \ \ \ \ \ \nonumber \\
\nu _0 &=& \ln \left( \frac{\left( 1-\frac{2B}{B+a}\right)
^{-\frac 1{A
\ }}\left( a-\mu \right) }{a+\mu }\right) , \ \ \ \ \ \ \ \ \ \ \nonumber \\
\phi _0 &=& \left( \frac{a+B}{a-B}\right) ^{\frac{2\left( a^2+\mu
^2-a\mu \right) }{A \ a\mu \ \omega }} .\ \ \ \ \ \ \ \ \ \ \ \ \
\ \ \ \ \ \ \ \
\end{eqnarray*}
The spatial distribution of the JBD scalar field is restricted to
the throat neighborhood, so that the dimensions of these wormholes
are not arbitrarily large. The junction surface, \textit{r} =
\textit{a}, is situated outside the event horizon, i.e.,
\textit{a}$> \mu $, to avoid a black hole solution.
\begin{figure}[!ht]
\centering
\includegraphics {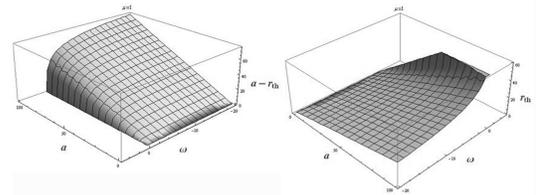}
\caption{The surface represents the value of throat of the
wormhole as a function of  $\omega $ and junction radius
\textit{a}} \label{plot1}
\end{figure}

Suppose we consider the Brans solution as a throat of wormhole. We
were led to this model by considering a bridge between separated
Schwarzschild and Brans Universes or a bridge connecting two
Schwarzschild asymptotically flat regions.
\begin{figure}[!ht]
\centering
\includegraphics {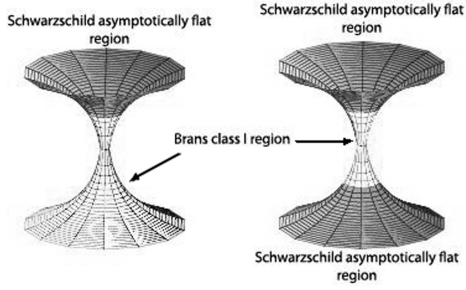}
\caption{Two types of the composite wormholes. The visualization
of the surface generated by the rotation of the embedded curve
about the vertical z axis. } \label{plot2}
\end{figure}

The same procedure can be easily adopted in other three classes of
Brans solutions.

\section{Other possible composite wormholes}
Finally, we should mention yet another proposal related to
wormholes. However, to our knowledge, exact complex scalar field
(boson stars) and for vector-metric wormhole solutions are
relatively scarce. The field equations of the vector-metric theory
\cite{Will} are obtained by the similar variational method as the
Einstein theory, and are given as following:

\begin{eqnarray}
R_{\mu \nu }-\frac 12Rg_{\mu \nu } + \omega \stackrel{%
\left( \omega \right)}{\Theta }_{\mu \nu} + \eta \stackrel{\left(
\eta \right) }{\Theta }_{\mu \nu} + \nonumber \\
+ \varepsilon \stackrel{\left( \varepsilon \right) }{\Theta }_{\mu
\nu } + \tau \stackrel{\left( \tau \right) }{\Theta }_{\mu \nu
}=8\pi T_{\mu \nu }^M, \label{eq202}\\
 \varepsilon K_{;\nu }^{\mu
\nu }+\frac 12\tau K_{;\nu }^{\mu \nu }-\frac 12\omega K^\mu
R-\frac 12\eta K^\nu R_\nu ^\mu =0.\nonumber
\end{eqnarray}
where $\stackrel{\rightarrow }{K}$ vector field and $\omega ,\eta
,\varepsilon ,\tau $ free parameters. Indeed, for the absence of
vector potential in empty space equations (\ref{eq202}) become
identical to field equation of Einstein theory. In a case of
static spherically symmetric spacetime we have Schwarzschild
metric
\begin{eqnarray}
&& g_{00}=-g_{11}^{-1}=1-\frac{2M}r\nonumber
\end{eqnarray}

Since Einstein and Bose it is well-known that scalar fields
represent identical particles which can occupy the same ground
state. Such a Bose-Einstein condensate has been experimentally
realized in 1995 for cold atoms of even number of electrons,
protons, and neutrons.

The Lagrangian density of Bose-Einstein gravitationally coupled
complex scalar field $\Phi $ reads

\begin{eqnarray}
L_{BS}=\frac{\sqrt{\left| g\right| }}2 ( R+[ g^{\mu \nu } \left(
\nabla _\mu \Phi ^{*}\right) \left( \nabla _\nu \Phi \right) - \nonumber \\
-U\left( \left| \Phi \right| ^2\right) ] ) . \label{eq212}
\end{eqnarray}

Variation of (\ref{eq212}) with respect to g$_{\mu \nu }$ and
$\Phi $ gives, the coupled Einstein-Klein-Gordon equations

\begin{eqnarray}
&& R_{\mu \nu }-\frac 12Rg_{\mu \nu }=T_{\mu \nu }^\Phi, \nonumber \\
&& \left(\nabla _\alpha \nabla ^\alpha + \frac{dU}{d\left| \Phi \right| ^2}\right) \Phi =0, \label{eq222}\\
&& \left(\nabla _\alpha \nabla ^\alpha + \frac{dU}{d\left| \Phi
\right| ^2}\right) \Phi ^{*}=0.\nonumber
\end{eqnarray}
where

\begin{eqnarray}
&& T_{\mu \nu }^\Phi =\frac 12\left[ \left( \nabla _\mu \Phi ^{*}\right) \left( \nabla _\nu \Phi \right) +
\left( \nabla _\mu \Phi \right) \left( \nabla _\nu \Phi ^{*}\right) \right] - \nonumber \\
&& \ \ \ \ \ \ \ \ -g_{\mu \nu }L\left( \Phi \right) \sqrt{\left|
g\right| }. \label{eq232}
\end{eqnarray}

 Obviously there is the solution of
field equation  in empty space don't depend on $\Phi $ and one can
assume the Schwarzschild metric as a vacuum solution for
Bose-Einstein complex scalar field.

Thus, it would be natural to expect that, similar to JBD theory,
one can construct composite wormhole solutions by matching an
interior wormhole spacetime to an exterior vacuum Schwarzschild
solution, in these cases too. We do not claim this list is
exhaustive.

\section{Conclusions}

We have constructed composite vacuum wormhole solutions by
matching an interior Brans solution to a vacuum exterior
Schwarzschild spacetime. To summarize the situation so far: The
technique developed in this note has helped us in several ways. It
led us to consider the static wormhole solutions in vacuum, where
the source of gravity being the scalar field. JBD theory can agree
with general relativity in empty space, it is important to study
the interior of wormhole in which the two theories may give
different predictions.

It furthermore contains interesting special case: a bridge between
separated Schwarzschild asymptotically flat region and region with
different spacetime geometry (Brans solutions or others). These
examples may not only provide further experimental and
observational tests that might distinguish between general
relativity and JBD theory, but they may also illuminate the
structure of both theories.

Wormholes require for their construction what is called "exotic
matter". Some classical fields can be conceived to play the role
of exotic matter. They are known to occur, for instance, in the
Visser's thin shell geometries \cite{Visser}, R+R2 theories
\cite{Hochberg}, Moffat's nonsymmetric theory \cite{Moffat},
Einstein-Gauss-Bonnet theory \cite{Gravanis} and, of course, in
JBD theory \cite{Camera}. Whichever it might be, it is very likely
that the same phenomena could also occur for complex scalar field
(boson stars) and vector field, too.

In scheme presented in this report, studies of possible wormhole
solutions in alternative gravitation was thought of as a way of
understanding the role of different fields as the "carrier" of
exoticity together with the aim of finding phenomena for which
different qualitative behaviors to those of standard General
Relativity model may arise.

\renewcommand{\refname}{References}

\end{document}